\documentclass{article}                     

\usepackage{graphicx}
\usepackage{bm}
\usepackage{amsmath}
\usepackage{amsfonts}
\begin{document}

\title{ Epistemological and ontological aspects of quantum theory}

\author{Inge S. Helland 
}

%\institute{Inge S. Helland \at
 %             Department of Mathematics, University of Oslo \\P.O. Box 1053 Blindern, N-0316 Oslo, Norway\\
%              Tel.: +47-93688918\\
%              \email{ingeh@math.uio.no}           
%}

\maketitle

\begin{abstract}

In this paper, epistemology and ontology of quantum states are discussed based on a completely new way of founding quantum theory. The fundamental notions are conceptual variables in the mind of an observer or in the joint minds of a group of observers. These conceptual variables are very often accessible, that is, it is possible to find values of the variables by doing experiments or by making measurements. An important notion is that of maximal accessibility. It is shown here that this new machinery may facilitate the discussion of when a specific quantum state can be given an ontological interpretation, and also the more speculative question whether \emph{all} states can be given such an interpretation. The discussion here is general, and has implications for the basic problem of how one should look upon information from experiments and measurement, in particularly the question concerning when this information may reflect properties of the real world.

\end{abstract}

\section{Introduction}
\label{intro}

The discussions on the foundation of and the interpretation of quantum mechanics have been in the literature since the theory was introduced more than 100 years ago. It has been particularly intense during the last decade, and to people outside the quantum foundation community, the situation has been confusing, to say the least. 

In his popular recent book [1] Lee Smolin discusses the realist position versus the anti-realist position. I will come back to that book below.

This paper, together with the book [2] and a few other recent papers referred to below, is an attempt to look at the situation with completely new eyes. In all other introductions to quantum mechanics and all research papers, the basic notions are unit vectors and self-adjoint operators in some separable Hilbert space. I want to consider these notions as derived notions. To me, the basic notions are conceptual variables existing in the mind of some observer or in the joint minds of a group of communicating observers. The state vectors, the operators and the Hilbert space are notions derived from this, as shown basically in Theorem 1 of the next Section. The discussion of the present paper gives one application of this theorem. Another illustration of the implications of the same theorem is in the discussion of the Bell theorem and the Bell experiment, as given in [3].

This theorem not only leads to a new interpretation, an interpretation generalizing the QBist interpretation. It also facilitates many discussions on fundamental questions. This is illustrated in the present paper by taking up the important  questions about whether or not or in what sense quantum mechanics should be seen as an epistemological theory or an ontological theory. To me, the epistemological aspect is fundamental. But an important question is when a specific state also can be characterized as an ontological state describing the real world, another, more speculative question may be if \emph{all} states in some sense can be given such an interpretation. 

I first review and critizise two recent papers on these questions that are using the traditional machinery. Then I give my own discussion using the notion of epistemic processes. I leave to the reader to judge which conceptual machinery is the simplest and most enlightening one.

The plan of the paper is as follows:  In Section 2 I formulate the basic theorem, proved both in [2], [4] and [5]. Section 3 takes up the notion of causality as discussed in the recent book by Judea Pearl and in some of the resulting statistical literature. Section 4 discusses briefly a valuable aspect of quantum theory based on traditional concepts: quantum picturalism. In section 5 I review a recent paper on the epistemology/ ontology problem that uses quantum pictorialism. Section 6 reviews another recent paper on the same problem. In Section 7 I introduce Convivial Solipsism, which as an interpretation is related to my own views, and in Section 8 I introduce measurements. Then in Section 9 I give my own discussion on when and how quantum states can be given an ontological interpretation. Two brief remarks are given in Section 10, before I  in Section 11 discuss briefly the recent book by Lee Smolin on the search for some reality beyond quantum theory and some related articles. I complete with some very general remarks in Section 12.

\section{Quantum theory from conceptual variables}
\label{sec2}

One main result from [2], as generalized in [4] and [5], is the following:
\bigskip

\textbf{Theorem 1} \textit{Consider a situation where there are two maximally accessible conceptual variables $\theta$ and $\xi$ in the mind of an observer or in the joint minds of a communicating group of observers. Make the following assumptions:}

\textit{ (i) On one of these variables, $\theta$, there can be defined group actions from a transitive group $G$ with a trivial isotropy group and with left invariant measure $\rho$ on the space $\Omega_\theta$.}

\textit{ (ii) There exists a unitary irreducible representation $U(\cdot )$ of the group $G$ defined on $\theta$ such that the coherent states $U(g) |\theta_0\rangle$ are in one-to-one correspondence with the values of $g\in G$ and hence with the values of $\theta$.}

\textit{ (iii) The two maximally accessible variables $\theta$ and $\xi$ can both be seen as functions of an inaccessible variable $\phi \in \Omega_\phi$. There is a transformation $k$ acting on $\Omega_\phi$ such that $\xi(\phi)=\theta(k\phi)$.}

\textit{Then there exists a Hilbert space $\mathcal{H}$ connected to the situation, and to every accessible conceptual variable there can be associated a unique symmetric operator on $\mathcal{H}$.}
\bigskip

Of course the Hilbert space $\mathcal{H}$ here is the one associated with the representation (ii) in the theorem. The most important result is that to every accessible conceptual variable there is associated a unique operator on this Hilbert space. Explicit formulas for the operators are given in [2], [4] and [5]. Note that this is for free without stating any abstract axioms.

To understand this theorem, some definitions are necessary: An accessible conceptual variable is a variable that one can in principle obtain values for by doing an experiment or performing a measurement. It is maximally accessible if it can not be taken as a function of another, wider accessible variable. A more precise mathematical definition of this is given in [5]: The conceptual variables have a partial ordering given by functional dependence: Say that $\theta$ is less than or equal to $\lambda$ if $\theta = f(\lambda)$ for some function $f$. Then by Zorn's lemma there exist maximally accessible conceptual variables.

  It is important that there in situations related to quantum theory as approached in [2], also exist inaccessible conceptual variables, like the full spin vector of a particle or the vector (position, momentum). Thus Heisenberg's uncertainly relation is an important background assumption behind the theorem, essentially the only physical assumption that is needed.

The assumption (ii) can be satisfied under weak conditions, as shown in [4] and [5]. In particular, it seems like this assumption can be omitted in the discrete case.

In the discrete case, more can be proved [2]: The possible values of an accessible variable coincide with the eigenvalues of the corresponding operator. The eigenspaces of the operators connected to variables $\lambda$ are in one-to-one correspondence with questions: `What will be the value of $\lambda$ if I measure it?', together with a sharp answer `$\lambda=u$'. If and only if $\lambda$ is maximally accessible, the eigenspaces are one-dimensional. This gives a concrete, very simple interpretation of very many unit vectors in the Hilbert space. Each such vector $|\psi\rangle$ can be said to be in one-to-one correspondence with a question-and-answer pair: `What is $\lambda$?' together with an answer `$\lambda=u$. The difficult problem of determining when \emph{all} relevant unit vectors in some concrete situation can have the above representation, is taken up in [6].

The assumption that there can be defined a transitive group acting upon $\theta$ is crucial. It can easily be satisfied when the range of $\theta$ is finite or is the whole line ${\mathbb{R}}^1$.

Theorem 1 can be used for a new foundation of essential parts of quantum mechanics. Having established this important foundation, the main other foundational result to prove, is the Born formula. In [2] this formula is proved under the following three assumptions: 1) The likelihood principle from statistics holds (this principle is motivated in Chapter 2 of the book). 2) The observer performing the relevant experiment or measurement has ideals which can be modelled by a perfectly rational concrete or abstract being. Rationality is here formulated in terms of a Dutch Book principle. 3) The state in the mind of this observer describing the physical system before the measurement or experiment, is given by a sharp value of some maximally accessible variable.

For further consequences of this theory, see [2].

\section{Causality, inference, and reality}
\label{sec3}

The book [2] is concentrating on epistemic processes, processes to obtain knowledge through experiments or measurements. (Of course, there are also other ways to obtain knowledge; this is largely ignored in [2].) A very important problem that remains to be discussed, is to what extent the results of such epistemic processes can be associated with some sort of reality, a `real' world. The only statement about this given in [2] is the following: If all real and imagined observators can be said to agree on the result of some experiment or measurement, then this is a strong argument to the effect that this result can be coupled to some reality. This conclusion is strengthened if the experiment is done in a `proper' scientific way.

Recently, a rather involved discussion of the reality-question has been done in  [7]. That discussion relies heavily on the notion of causality, so our first task here will be to review recent scientific developments that to a large extent has clarified understanding and use of the causality concept.

The book by Pearl [8] has been extremely important in clarifying the distinction between causality and statistical inference. Very briefly, statistics has to do with observing data, postulating a model describing these data, and drawing conclusions from that. As an example, a phenomenon under investigation may be the correlation between different variables. On the other hand, any causal statement involves a direction between variables: An assumed cause and an assumed effect. The main point in [8] is the introduction of a `do' statement, and an intervention related to this: If I assume that I can do something with one variable (the cause), then this may have an effect upon another variable, the effect or response.

Diagrams are important in [8], diagrams involving arrows (cause-effect relations) from one variable to an other. These diagrams can be quite complicated, but are always useful for trying to understand causality. In [9] Cox and Wermuth used diagrams together with probability models to explain the difference between conditioning and intervention. Three variables are involved: Cause (C), response (R) and background (B). In the physical litterature, and also in [2], the background variable may be taken as synonymous to the context of the situation. In general there is an arrow from B to R, an arrow from B to C, and an arrow from C to R. In probability terms this can be written in the recursive form
\begin{equation}
f_{RCB}=f_{R|CB}f_{C|B}f_{B}.
\label{1}
\end{equation}

Conditioning i Pearl's sense is a standard conditioning calculation in (\ref{1}) given that C is fixed at some value c. The resulting conditional distrubution of R can be found by marginalizing over B, that is
\begin{equation}
f_{R|C}=\int f_{R|CB}f_{B|C} db,
\label{2}
\end{equation}
where $f_{B|C}=f_{CB}/f_{C}$.
According to [9] it is appropriate ro use $f_{R|C}$ for constructing an empirical prediction of $R$ given only $C=c$.

To represent an innovation, on the other hand, one must make an assumption about the causal connection: In terms of the diagram described above, the arrow directly between $B$ and $C$ must be removed, corresponding to an assumption of independence. This in general implies a different distribution of $R$, when we have intervened to set $C=c$, namely
\begin{equation}
f_{R||C} = \int f_{R|CB}f_{B} bd.
\label{3}
\end{equation}
This corresponds to Pearl's definition of a causal effect.

The probabilities in (\ref{1})-(\ref{3}) may in principle be any probabilities, for instance quantum probabilities.

In [2] the operators and the quantum states are coupled to questions to nature of the form `What will the conceptual variable $\theta$ be if we measure it?', that is, a `What?- question. In [8] causal diagrams and their realization are also coupled to questions of the type `How?' and `Why?' According to the philosopher David Hume, the why is superflous as it is subsumed to the how. The last chapter in [8] discusses various philosopher's opinions on the causality concept.

In any case, when discussing reality in quantum mechanics, Schmid et al. [7] indicate a theory where they look at both causality and inference in order to enlighten this discussion. Before I enter a brief review of that paper, I have to go into the recently developed diagrams describing quantum theoretical processes.

\section{Quantum picturalism}
\label{sec4}

The following is a very brief recapitulation of a deep and important subject.

[10] is a review paper that builds upon deep mathematics (category theory, specifically monoidal categories), but at the same time illustrates the relevant theory by almost trivial, everyday sets of processes. Diagrams are described as tools for intuitive reasoning about complicated interacting quantum systems, a step-stone towards a deeper conceptual understanding of quantum theory. Elementary quantum mechanics is seen to be `low level'; the diagrams allow one to reason at a much higher level.

Basically the relevant diagrams consist of boxes bound together by one or several wires. Algebraic structures defined in terms of elements, operations and axioms, are in one-to-one correspondence with pictures. Any equation between two pictures is derivable from intuitive rules in the diagrammic calculus, if and only if the corresponding formal expressions are derivable from the axioms of the algebraic structure.

In a quantum mechanical setting, the boxes are said to represent processes, the wires systems. States are special processes without any input wire; effects are special processes without any output wire. A wire (system) may be one of  the results of some process, or it may be  one of the factors that initiate a process.

I want to add an additional remark to my own notion of a state. Conceptually, a state may be defined through a choice between several possible conditions that characterizes a given  physical system. In concrete terms, in [2] a state is associated with some relevant maximally accessible conceptual variable and a sharp value of this variable, a specific set of conditions. Thus from my point of view, what is often not stressed in physics texts, seemingly also not in [10], is that the specific set of conditions must either be chosen initially by some observer or by a communicating group of observers, or chosen during observation by an observer/ group of observers. In this sense the states can be said to belong to certain mind(s) of observer(s). In reality, a whole diagram, in the way it is described in [10] must be seen as a model, belonging to the mind(s) of some observer(s). I will come back to this later.

\section{Starting to unscramble the omelette}
\label{sec5}

The analysis of [7] takes as a point of departure a statement by E.T Jaynes to the effect that our present quantum mechanical formalism is in part describing realities of Nature and in part incomplete human information about Nature. This is scrambled into an omelette which seems to be difficult to unscramble. The goal of the authors of [7] is to try to approach this by looking partly upon causal mechanisms and partly on inferential mechanisms in the relevant sets of processes, and developing a theory that takes both these aspects into consideration.

An important notion in [7] is that of a diagram-preserving map. Diagrams (pictures) are defined as in Section 4 above, in particular the processes are the boxes of the diagrams (with wires connected to them). A process theory is defined as a collection of processes, $T$, which is closed under forming of diagrams. A diagram-preserving map, $\bm{m}: T\rightarrow T'$ is a map from prcesses in $T$ to processes in $T'$ such that wiring together processes in $T$ to form a diagram and then applying the map $\bm{m}$ to each of the component processes and finally wiring them together in $T'$.

Causal systems and inferential systems are defined in terms of diagrams. I will not go into details here, but it is important that in [7], inference is limited to inference by means of Bayesian probability theory. Two kinds of causal theories are considered. Causal-inferential theories are defined formally by an injective map from the causal theories and another injective map from the inferential theories. This is done in the classical case and in the quantum case. Inferential equivalence is then defined.

All these definitions allow the authors of [7] to describe how to `make an omelette', and from this classical realism is first discussed, and then how to go beyond classical realism. During this discussion, an important principle is introduced, claimed to have been endorsed by Einstein, what is called Leibniz's methodological principle: If an ontological theory implies the existence of two scenarios that are empirically indistinguishable in principle but ontologically distinct, then the ontological theory should be rejected and replaced with one relative to which the two scenarios are ontologically identical.

[7] lists several criteria that must be satisfied in order that some causal theory should contain enough structure to be doomed worthy of the title `realist'. A nonclassical realist representation map is said to be Leibnizian if it preserves inferential equivalence relations. An aim of the work is said to generate a compelling interpretation of quantum theory - one that satisfies the spirit of locality and Leibnizianity. During the discussion it is among other things said that the dichotomy in the interpretation of quantum theory: Does a state have an ontological or epistemological status? - sometimes is a false dichotomy. The discussion in [7] aims at subsuming both classical and non-classical causal modelling. The framework in the quantum case seeks to circumvent certain no-go results. It is argued that if one can identify their conventional (classical) interpretations, then one has the means of defining an  intrinsically quantum notion of realism. The discussion is not assumed to be final; several possible extensions are sketched.

\section{Shall we give up classical ontology?}
\label{sec6}

No one should seriously doubt that the real world exists. Questions like `Is the moon there when nobody looks?' [11] may be asked, but not really seriously. In my opinion, the question of realism should not be concerned with the \emph{existence} of the real world, but with the detailed joint existence of all \emph{properties} of the real world.

However, the recent paper by Evans [12] takes as his point of departure what is called Einstein-Bell realism. It starts with 3 conditions:

1) Quantum mechanical probabilities are epistemic;

2) Quantum mechanics is local;

3) Quantum mechanics is consistent with the no-go theorems.

Einstein-Bell realism is the conjunction of these conditions with the assumption that there is an objective reality.

The no-go theorems play an important role. Three are discussed in [12]: (i) The Bell theorem; (ii) The Kochen-Specker theorem; and (iii) The recent Shrapnel-Costa theorem. The Shrapnel-Costa theorem is seen as an extension of the Kochen-Specker theorem. Briefly it states that any ontological underpinning quantum behaviour must be contextual; moreover, ``what is contextual is not just the traditional notion of `state', but any supposedly objective feature of the theory, such as a dynamical law or boundary condition, which is responsible for the experimentally observed statistics.''

Evans considers what he calls the causal symmetric local hidden variable approach to interpreting quantum theory as the most natural interpretation that follows from Einstein-Bell realism. As we know, there exist very many interpretations of quantum mechanics. A strange feature of the causal symmetric local hidden variable approach is that it seems in a way to go against the natural time ordering when discussing causality.

Anyway, the conclusion of [12] is as follows: Causal symmetry is incapable of explaining quantum behaviour as arising as a result of noncontextual ontological properties of the world. This leads, Evans says, to a rejection of Einstein-Bell realism. More than this, it is argued that, even where there seems to be a possibility of accounting for contextual ontic variables within a causally symmetric framework, the cost of such an account undermines a key advantage of causal symmetry; that accepting causal symmetry is more economical than rejecting a classical ontology. Either way, he says, it looks like we should give up on classical ontology. 

\section{Convivial Solipsism}
\label{sec7}

The philosophy of convivial solipsism in the context of quantum measurements is discussed in detail in Zwirn [13, 14].  A brief recapitulation will be given here.

In general, solipsism is a philosophy with many variants. It is based upon the view that everything that we can know for sure by our mind is connected to this mind. My mind is an autonomous separate world. The convivial variant also recognizes that other people have their minds and thus have sure statements connected to their minds. And communication between different people is possible. In my opinion, people that have communicated and agreed on certain questions, may be seen as a new unit with respect to these questions, in the sense that they then may make common decisions.

In the microscopic setting defined by the EPR experiment and the Bell experiment, one can look upon Alice and Bob from the point of view of convivial solopsism. For each observer Zwirn's convivial solopsism is built upon two assumptions: 1) The quantum states that are meaningful to the observer are relative, not absolute. This assumption goes back to the writings of Everett [15], and can also be found in the QBism interpretation (Fuchs [16]) and in Rovelli's relational quantum mechanics (Rovelli [17]). 2) There is no measurement until the result of the measurement is perceived in the consciousness of an observer, and for this result there is a hanging-on mechanism: All subsequent measurements must be seen from the basis of this result.

Convivial solipsism can be coupled to a theory of decisions; in a quantum setting, see for instance Yukalov and Sornette [18]. Decisions can be made by a single observer, or they can be made by groups of observers, after there has been some communication between them. Thus communication is important. In the book [2], communication is seen upon as based on conceptual variables, variables which may have their origin in a single mind, or is the combined minds of several, communicating people. By the result of Theorem 1 above, these variables, and symmetry groups defined on them, are the points of departure for a logical development ending with an essential part of the formalism of quantum theory.

Zwirn's version of convivial solopsism is directly based upon the quantum formalism; my version is based simply upon conceptual variables. In my opinion, these variables occupy a large place in the minds of people, and in the combined minds of communicating groups of people. Having said that, however, I agree with the assumptions 1) and  2) above made by Herv\'{e} Zwirn. Concerning the measurement as received by the mind of an observer, and then conserved with respect to future measurements, I look upon this in many cases as a result of some statistical inference.

\section{On measurements}
\label{sec8}

Imagine an observer Alice who is planning to do some physical measurement. In her mind she has several physical variables connected to the actual context, and she wants to concentrate on one of them, say $\theta$. For simplicity assume that $\theta$ in this context and for Alice, is what I have called maximally accessible, and assume also that it takes a finite number of values $u_1,..., u_d$.

The following is important to me: $\theta$ is a variable connected to the physical setting that Alice wants to address during her measurement, but as she plans this measurement, it must also be said to exist in her mind. This double existence is crucial for my own approach towards quantum theory.

Now in this situation let Alice and her physical environment be observed by another observer Charlie. Charlie also has in his mind the actual variable $\theta$. He is not doing any measurement, but he looks forward to receive a report from Alice after she has done her measurement. Also, he is in a better situation than Alice, and for him, $\theta$ is not maximally accessible. He can also imagine other variables $\xi, \lambda, \rho,...$ which Alice could have measured. Assume that he knows that one of these, say $\xi$, also is maximally accessible for Alice. Define $\phi$ as being the vector $(\theta, \xi, \lambda, \rho,...)$, which varies in some space $\Omega_\phi$. Then $\theta$ is a function of $\phi$: $\theta=f(\phi)$. Assume that $\theta$ and $\xi$ take the same number of values (or both vary on the real line). Then let $k$ be the transformation on $\Omega_\phi$ which switches the first two coordinates, so that $\xi=f(k\phi)$. \emph{Then the essential conditions of Theorem 1 above are satisfied if we look upon the situation from Alice's point of view}. In the finite case, the transitive group $G$ on the range of $\theta$ can be defined as generated by the transformation $u_1\rightarrow u_2, u_2 \rightarrow u_3,...,u_d\rightarrow u_1$.

\section{The epistemic process approach}
\label{sec9}

\subsection{Conceptual variables}
\label{subsec9.1}

What is my opinion about the different approaches to the question of ontology sketched above? First, the analysis in [7] is involved and impressive, but I disagree with its conclusions for two reasons: First, I see inference as being more general than its Bayesian version. Secondly, I cannot see that ontology should be directly coupled to the questions related to cause and effect. Considering the paper by Evans [12], I consider it to be enlightening to discuss whether or not one should give up classical ontology. I will give my own analysis below in terms of conceptual variables. Everything that has been written in this paper until now, must be seen as a background for this discussion.

In a concrete physical measurement setting, quantum mechanics has as its input certain statements and as its output some other statements. In reality such inputs and outputs characterize any empirical scientific investigation.  In many cases these statements in connection to measurements and experiments are obviously statements related to the real world. This may be true for any reasonable empirical scientific investigation. In a physical setting it may be statements about physical variables, variables that are measured. In my theory \emph{the variables also must be said to have an existence in the mind of an observer or in the joint minds of a group of communicating observers.} This way of seeing things is inspired by statistical theory. Statistics relies on models of reality, models of data in terms of parameters. These statistical parameters are special cases of what I call conceptual variables. They exist basically only in the minds of the researchers that analyse the situation.

To give a very simple relevant macroscopic example, let us assume that we are given some object A, and ask 'What is the weight of object A?'. Then $\mu=$'weight of A' is an physical variable. \emph{But both before and after and during the measurement it also exists in minds of us as measurers.} We can use a scale to obtain a very accurate estimate of $\mu$. Or we can use several independent measurements, and use the mean of those as a more accurate estimate. In the latter case it is common to introduce a statistical model where $\mu$ is a parameter of that model. But in my view the conceptual variable notion is more fundamental. The variable $\mu$ exists before any statistical model is introduced. I think that many people will agree that $\mu$ also during measurement exists in some sense in the mind of the measurer(s). In this example the conceptual variable has some ontic basis, but I will claim that this need not always be the case in all epistemic processes. Even in this case the existence of $\mu$ as a real number may be discussed. For instance, the question 'Is $\mu$ rational or irrational?' is rather meaningless. 

I will give two simple examples of conceptual variables that do not necessary have an ontological interpretation, one macroscopic and one microscopic.

1. Assume that Bob has two apples, and says to Alice that when he sits, he will have either one or two apples in his lap. Then he hides behind a screen and asks Alice: `Do I have one or two apples in my lap?' Alice tries to guess, and during this game $\theta=$ `the number of apples in Bob's lap' is a conceptual variable in Alice's mind. Finally, Bob discloses that he is standing, so $\theta$ is meaningless, has no connection to the real world.

2. Assume that Alice is doing some quantum measurement. Her initial mental state related to physical system that she wants to measure, is $|\psi\rangle$, corresponding to $\theta=u$ for some maximally accessible conceptual variable $\theta$. She wants to measure another maximally conceptual variable $\xi$, and after the measurement she will have a new state related to this variable. But what is her mental state \emph{now}, during the measurement itself? At that point of time neither $\theta$ nor $\xi$ take any value; in particular, it is certain that at this point of time neither of them have any ontological interpretation.

It is crucial now to analyse further the meaning of conceptual variables connected to an observer or a group of observers, in particular in connection to quantum measurements.

\subsection{Measurement, observer, and state}
\label{subsec9.2}

So assume that Alice is doing some measurement in a concrete context, which also may depend upon settings chosen by herself. Again let her initial state be $|\psi\rangle$, corresponding to $\theta=u$ for some maximally accessible variable $\theta$. I can generalize to unknown pure states, in particular mixed states given by a density operator, but I want to keep things simple. I will assume that $\theta$ takes the values $u_1, u_2,..., u_d$; $u$ is one of these values. This pure state represents the relevant knowledge that Alice has before the measurement. After the measurement is done, her knowledge changes, say to $\xi=v$ for another maximally accessible variable $\xi$. Since $\xi$ is maximally accessible, she must now reject her previous knowledge. The question is whether or not the new knowledge can be seen as objective knowledge about the real world, ontological knowledge. I will come back to this question.

Now let the physical system together with Alice be observed by another observer Charlie. In particular, he may have the possibility of knowing the state of Alice both before and after the measurement, but he may have a wider horizon, so that he also can have other knowledge. This means that his Hilbert space in the situation is larger than the Hilbert space describing the mind of Alice. Look at his state before he notes the particular value of $\theta$, although he focuses upon Alice's possible value for this variable. This can be written
\begin{equation}
|\phi\rangle = \sum_i c_i |\psi_i\rangle |\rho_i\rangle + |\phi_0\rangle ,
\label{5}
\end{equation}
where $|\psi_i\rangle$ represents the result $\theta=u_i$, $|\rho_i\rangle$ represents other related knowledge in the situation, and $|\phi_0\rangle$ is a state orthogonal to all the $|\psi_i\rangle |\rho_i\rangle$.
\bigskip

As a simplifying assumption now, assume that $|\phi\rangle$ is a state representing a concrete knowledge in the mind of the relevant observer Charlie, say $\alpha =w$ for some maximally accessible variable $\alpha$. An example of an entangled state with such an interpretation is given in [3].

One possible scenarium is that Charlie at some pont of time shares his knowledge with Alice. Acording to Convivial Solipsism this must be treated as a measurement done by Alice. Her state resulting from this will then be a relative state of the type introduced by Everett [15]. Specifically, the entangled joint state describing the mind of Alice and the mind of Charlie is
\begin{equation}
|\phi_1\rangle = \sum_i c_i |\psi_i\rangle |\lambda_i\rangle|\rho_i\rangle + |\phi_{10}\rangle ,
\label{7}
\end{equation}
where $|\lambda_i\rangle$ is the state of mind of Alice when $\theta=u_i$, and $|\phi_{10}\rangle$ again is orthogonal to the rest. According to [15] it is meaningless in this context to ask of the absolute state of the subsystem `Alice's mind', one can only talk about the relative state, given the result of the measurement as perceived by Charlie.

\subsection{Several observers}
\label{subsec9.3}

Charlie may also share his knowledge with other observers David, Elena,..... These observers then form a communicating group of observers. As a group they may also have other knowledge, but their joint state always includes the specific knowledge $\alpha=w$. This will most probably have implications for joint decisions made by this group.

Of course a similar discussion can be made if Charlie focuses on the state of Alice \emph{after} she has done her measurement. It is an important observation that a quantum state may depend on the focus chosen by an observer.

Let us go one step further. Assume that Charlie and his group is observed by another observer George, having a still larger Hilbert space in the situation. George may also be interested in (focus on) other physical situations and other groups of observers. Making simplifying assumptions as above, his state may be specified as $\beta=z$, where $\beta$ is a new conceptual variable, maximally accessible for George. If George again should share his state with Charlie, or perhaps with Alice, this knowledge will give these observers new relative states.

And one can continue: George is observed by another observer with larger resources and so on. 

The discussion above was limited to quantum mechanics. But note that in all cases of the above discussion, only a state of the form $\lambda=z$ was assumed in connection to the relevant observer. Hence the discussion below is relevant in connection to any person seeking some knowledge, as long as this knowledge-seeking can be expressed in the form of a question `What will be the value of $\lambda$ when I try to find this value?' together with a sharp answer $\lambda=z$. In connection to ordinary measurements, where no other complementary other related maximally accessible variable is involved, the Hilbert space must be seen as one-dimensional, and ordinary statistical theory applies to the situation.

But what has all this to do with the possible ontology of states? Here is my opinion: Assume that we stop the observation process above at some point, say with George and his conceptual variable $\beta$. Then the nature of this conceptual variable is crucial. George has to make a decision based upon the situation: Is it natural to regard $\beta$ as some property of the real world? This type of decision is by its nature something that is \emph{completely outside the formulation of the quantum mechanical model or other relevant mathematical models}, the former derived from assumptions around Theorem 1 as described above. Thus the ontology question is something that relies on independent decisions, but is coupled to a defined conceptual variable, and how it is related to the physical or other concrete situation, as perceived by an observer.

\subsection{Alice's possible views on the reality question}
\label{subsec9.4}

The simplest situation is just when we stop the above process with Alice and her measurement situation. Alice has to decide whether or not, as she see it, her conceptual variable $\theta$ is a property of the real world. In a physical context, say, if $\theta$ represents the charge or rest mass of a particle, she will probably say yes. In other cases she might be in doubt. Note that according to Convivial Solipsism, every property of the world must be related to the mind of some observer, here Alice. So also the question whether or not a conceptual variable can be seen as a property of the outside world; this question must be decided by her.

The decision here taken by Alice, will depend on her philosophy and on the nature of the conceptual variable $\theta$. For instance, $\theta$ may be a property of a concrete particle focused upon after an event at the LHC at Cern. If $\theta$ is the charge or rest mass of the particle, it is obvious that Alice will look upon it as some property of the real world. But assume that $\theta$ is the position of the particle at time $t$ relative some frame defined by Alice, or its velocity at time t relative to this frame, or the time that it reaches some given $x$-value. Then Alice's philosophy is crucial. She may be a simple realist, insisting, quite independently of whether quantum mechanics is true or not, or a complete theory or not, that every measurement should have some immediate relation to some objective reality. More elaborate versions of this view can be attributed to Albert Einstein, whose views were recently taken in a more modern context by Smolin [1]. 

To Smolin, quantum mechanics is not the final story. He states in [1] that there is still much to discover, and in several recent papers he has tried to elaborate on this. I will come back with some brief remarks to these papers in Section 11.

There seems to be a basic difficulty with the simple version of the realist view, if we at the same time should accept the concept of complementarity, a concept that in my opinion should lie behind any reasonable theory of the world. Say that Alice has the choice between measuring two maximally accessible variables $\theta$ and $\xi$, but that she actually measures $\theta$. Then, by the simple version of realism, only $\theta$ has a real existence, and thus the choice made by Alice, seems to, after this view, be the only possible choice, implied in some way by her context and her history. This may be called superdeterminism, and it limits in an essential way Alice's free will. 

As is well known, an anti-realist view was taken by Niels Bohr. His views are founded directly on the concept of complementarity, a concept that I in [2] extend also to many macroscopic situations. Then, for example, if Alice measures some position variable, she must also have in mind that she \emph{could} have measured the complementary variable, velocity. 

 In general, I interpret the anti-realist view to imply that one also believes in the possibility of a free will. If this also is Alice's opinion, it will influence her view of reality. Since by her free choice she can focus on either position or velocity, and since by complementarity only one of these are allowed, she cannot possibly take both position and velocity to be properties of the real world. 

Other, related, views of the reality question in connection to measurements have been given for instance in [16] and [17]. The debate on these issues seems to be hard now, and is confusing to anybody outside the quantum foundation community. My own opinions are strongly influenced by Niels Bohr.

The statement that we cannot take both position and velicity to exist in the real world, must be taken with one qualification: For macroscopic objects, both exist `for practical purposes'. 

But my point here is the following: Our views on reality in connection to measurement depend on 1) what conceptual variable we focus upon, 2) what philosophy we have in connection to quantum measurements. I claim that both 1) and 2) are independent of quantum theory as seen as a mathematical framework for modelling and prediction. This theory may or may not exist in the mind of Alice. Complementarity is a basic property behind any reasonable theory, in particular quantum mechanics, but acceptance of the detailed quantum theory is another issue. Alice may not be aware of Theorem 1. 

\subsection{The views of the other observers, and some speculations}
\label{subsec9.5}

A similar discussion can be carried out for Charle, George and all other observers described in the process above.
An  extreme is that we continue this process until it covers all human beings in the world. Of course we then enter into what must be called pure speculations.

But here is part of my opinion: One may believe that all possible people and all physical situations are observed by a superior observer, which one can call God. This is of course only one possible belief, far from shared by all people. But if one should have such a belief, one can use arguments like the ones above to look at the states as perceived by this God, in the way described here. All these states can be related to some big entangled state descibing the situation from the point of view of God. If one takes the view of a realist, they may be even seen as ontological states. Taken to the extreme, this may seem to imply superdeterminism and the lack of free will, perhaps a not too attractive solution.

But note that, if there is a God, he does not in any way seem to share his complete state with human observers. Various religions and different religious communities have different views of God and of what one can say about nature related to Him. These are complementary views.

Also, if we believe in a God which at the same time is almighty and perfectly good, we have a real difficulty with understanding all the suffering in this world. As I see it, this issue should be connected to the complementarity concept again, and to the assumption that all humans have a free will. If God should do the absolutely best for all of us, He by necessity is faced with many complementary decisions, so by pure logic it is impossible for Him to satify us all. 

Thus, as I see it, the question of whether or not one should believe in the interpretation of all quantum states as ontological states, seems to be related to the question of a higher abstract being related to this world. This calls for many interesting further discussions. My own views of science and religion are elaborated in the papers [19, 20].

\subsection{Some concluding remarks}
\label{susect9.6}

Thus, the concept of a higher abstract being can be used in connection to possible real knowledge about every person, and one could also say in connection to all properties of the world. A far less exotic situation will be if the physical setting is such that the variable $\theta$ must be looked upon by everybody as a real physical variable, and arbitrary groups of observers naturally focus on $\theta$ and get the same value for this variable. Again charge and rest mass may be examples, on the other hand, according to relativity theory, space intervals, energy and momentum variables are not objective in this sense. In these cases, also from the point of view of quantum mechanics, extra asumptions of the kind used in the above discussion are needed.

But in the case of charge, for instance, everybody will agree that $|\psi\rangle$ corresponding to some $\theta =u$ must necessarily be an ontic state, connected to the real world. As seen from quantum mechanics, such states are trivial; the Hilbert space is one-dimensional. So also in other knowledge-seeking cases where no complementarity is involved.

 This whole discussion relies on the assumption that some variant of Theorem 1 is true, and that states, as discussed in [2] can be characterized by question-and-answer pairs in terms of maximally accessible conceptual variables.

The discussion above holds for arbitrary pure states, for instance the states before or after a unitary state transformation, a transformation resulting automatically from the Schr\"{o}dinger equation or consciously from manipulations by an observer (see for instance Hardy [21]).

A large range of conceptual variables may be used in this discussion, for instance variables defined in term of complex models, like the pictorial models of Section 4 above.

\section{Two additional points}
\label{sec10}

It strikes one how ubiquitous [22] quantum theory is. Not only does it give a rather complete theory of the microcosmos, recently it has also been employed to cognitive models [23], to finance [22], to basic decision theory [18, 23] and to the study of human language [24]. In this paper I have concentrated on its physical applications. However, some of my remarks are also relevant to these other application fields.

In physical applications the process of obtaining knowledge through experiments and measurements is important. In general, increased knowledge may make it possible to make better decisions, although this is not always true, as indicated in a particular very simple statistical situation in [25].

\section{A particular recent view upon the reality question}
\label{sec11}

The debate on reality started by Albert Einstein and Niels Bohr has continued to these days. Lee Smolin in his book [1] has modernized Einsteins views, and declares himself as a realist. To do so, he answers yes to each of the following two questions:
\bigskip

1) Does the natural world exist independently of our minds? More precisely, does matter have a stable set of properties in and of itself, without regard to our perceptions and knowledge.
\bigskip

2) Can those properties be comprehended and described by us? Can we understand enough about the laws of nature to explain the history of our universe and predict its future?
\bigskip 

Myself, I will obviously answer yes to the first part of question 1). To the second part, I am more in doubt. If one only talks about some of the properties, the answer is again yes. If one should talk about \emph{all} properties, then according to the discussion in Section 9, the answer depends on our philosophy. As I see it, a simple realist position may lead to superdeterminism as a necessary assumption.

Then again, to question 2) I will only say a conditional yes. We can comprehend and describe many properties, but not all. We are limited, a general conclusion also made in [3]. In particular, we can predict the future to some extent, but not completely.

Taking his yes to the same two questions as a point of departure, Smolin [1] has an interesting and long discussion, during which he also describes the views of many other researchers and of many other schools in relation to quantum interpretation. His own final views are formulated in the epilogue. There he claims that a radically new theory is needed to solve the fundamental issues in physics and cosmology. And to this end, he says that he do not even know how to begin.

Smolin has now developed his views in several articles. I will concentrate on one recent such article [32], which is based upon the articles [33] and [34]. Together, these articles give a deep new theory, which I cannot go into details with here. Fundamental notions are those of events. In [32], views are attributes of events, and the theory's only beables; they comprise information about energy and momentum transferred to an event from its causal part. Quantum mechanics is derived from the principle that the universe contains as much variety as possible [34], in the sense of maximizing the distictiveness of each subsystem. An ensemble interpretation [33] is connected to the real ensemble of all systems in the same state in the universe. In [32] a dynamics is proposed for a universe constituted of views of events. It is stated in the conclusion that a great deal more remains to be done to develop the proposals.

I do not see my own approach as giving a radically new theory. To me, quantum theory is a valid theory describing the microscopic world, and also having applications in the macoscopic world, such that in Quantum Decision theory and in certain cognitive models. I only claim that a simpler and more intuitive basis for the theory can be given, as described in [2] and summarised in Section 2 above.

According to the discussion of Section 9, a complete ontological model of the world can be coupled to the assumption of a perfect and completely realistic higher being describing everything. In such a model, superdeterminism may seem to hold, and we have no free will.

I absolutely prefer a view of the world where we all may be limited in our decisions, where we have a free will, and are responsible for our actions. In my opinion, Niels Bohr's concept of complementarity will be important in such a world; see the last chapter of [2].

\section{Final remarks}
\label{sec12}

The purpose of this paper together with related papers and the book [2] has been to sketch a new, and in my opinion quite promising, attempt to understand quantum theory, causality theory and statistical inference theory from a common basis. Empirically, all these theories have been verified to an impressive degree, but the interpretation of quantum theory has been the source of much confusion. The book [2] is an attempt to develop the epistemic side of a new foundation, and from this propose a new interpretation: In every application, but also more generally, every state is connected to the mind of a single person or to the joint minds of a group of communicating persons. In this mind/ these minds conceptual variables exist, and these make the basis for results like Theorem 1 above.

One important background for the development of [2] has been that, in my opinion there has been too little communication between researchers working with the foundation of quantum theory and researchers in the statistics society. My book has been an attempt to develop elements of a future common culture.

What is culture? According to the author and philosopher Ralph D. Stacey it is a set of attitudes, opinions and convictions that a group of people share, about how one should act towards each other, how things should be evaluated and done, which questions that are important and answers that may be accepted. The most important elements in a culture are unconscious, and cannot be forced upon one from the outside.

In my opinion it is very important in the long run to try to develop a common international cultural foundation for all scientists. In this way, science could really achieve the authority that it requres and needs. There are extremely many problems facing the world now: climate, health problems, poverty problems, refugee problems, international conflicts and the existence of very dangerous weapons. For all these problems one should try to device rational solutions that at the same time satify good ethical standards; in particular, rational and ethical decisions should be made by national and international leaders. Unfortunately, in the processes of making decisions, we may all be limited. (In a concrete quantum situation, this has bee argued for in [3]). So, ideally, to arrive at good decisions, in addition to other requirements, our leaders should try to get  rational inputs. Ideally again, such inputs could in each concrete case be given by joint efforts by groups of scientists.

In improving this situation, we may all be somewhat responsible. In particular in this connection, the process of achieving knowledge is important. And, again in my view, such a process can always be associated with the mind of a single person or by the joint minds of a group of communicating persons. Knowledge as such may or may not be coupled directly to an external objective real world.

In his first book [26] Lee Smolin has listed 5 great problems in theoretical physics. Combining the conclusions of [2], [3], [4] and the present article seems to me to be a way to device at least a tentative solution to a large part of problem 2:  Resolve the difficulties in the foundation of quantum mechanics. This is done partly through appealing to a particular unification: Try to find a  unified new culture that is meaningful both to physicists and to statisticians. As I see it, this may also perhaps be a stepping stone for approaching further problems, like finding connections between elementary quantum theory and other theories, in particular field theory, theories in particle physics and relativity theory.

To find a conceptual basis from which quantum theory and general relativity both can be understood, is one of the most challenging problems in modern physics. Many researchers and several research groups have made their proposals on how to attack this problem. The most well known approaches are the following three: 1) Quantum loop theory. (For a popular partial account, see Rovelli [27]). 2) String theory. (For a brief introduction, see Susskind and Lindesay [28]). 3) The pure mathematical modelling approach. (See for instance Laudal [29]). From my point of view the operational approach by Hardy [30] may be particularly enlightening. More relevant references can be found in the latter paper. In my potential own approach to this problem I would consider looking upon space and time and other variables as conceptual variables, inaccessible inside black holes. Note that such a discussion will be independent of advanced concepts like loops or strings.

Concerning quantum field theory and the standard model, a good treatise is given in [31]. Group theory and group representation theory play a large role here. Following my ideas as sketched above, it might be fruitful to try to relate the abstract groups that lies behind the standard theory to group actions on some maximally accessible conceptual variables of physical relevance. But such investigations belong to the future.

\section*{References}

1. Smolin, L.: Einstein's Unfinished Revolution. The Search for What Lies Beyond the Quantum. Penguin Books (2019)

2. Helland, I.S.: Epistemic Processes. A Basis for Statistics and Quantum Theory. Second Edition. Springer, Cham, Switzerland (2021)

3. Helland, I.S.: The Bell experiment and the limitation of actors. Foundations of Physics 52, 55 (2022)

4. Helland, I.S.: On reconstruction of the Hilbert space from conceptual variables. arXiv: 2108.12168 [quant-ph] (2021)

5. Helland, I.S.: On reconstructing parts of quantum theory from two related maximal conceptual variables. International Journal of Theoretical Physics 61, 69 (2022)

6. Helland, I.S.: When is a set of questions to nature together with sharp answers to those questions in one-to-one correspondence with a set of quantum states? arXiv: 1909.08834 [quant-ph] (2019)

7. Schmid, D., Selby, J.H. and Spekkens, R.W.: Unscrambling the omelette of causation and inference: The framework of causal-inferential theries. arXiv: 2009.03297 [quant-ph] (2021)

8. Pearl, J.: Causality. Second Edition. Cambridge University Press, Cambridge (2009)

9. Cox, D.R. and Wermuth, N.: Causality: a statistical view. International Statistical Review 72 (3) 285-305 (2004)

10. Coecke, B.: Quantum picturalism. Contemporary Physics 51 (1) 59-83 (2010)

11. Mermin, N.D.: Is the moon there when nobody looks? Physics Today 38, 38-47 (1985)

12. Evans, P.W.: The end of a classical ontology for quantum mechanics? Preprint (2021)

13. Zwirn, H.: The measurement problem: Decoherence and convivial solipsism. Foundations of Physics 46,635-667 (2016)

14. Zwirn, H. Nonlocality versus modified realism. Foundations of Physics 50, 1-26 (2020)

15. Everett, H.: Relative state formulation of quantum mechanics. Reviews of Modern Physics 29, 453-462 (1957)

16. Fuch, C.A. : QBism, the perimeter of quantum Bayesianism. arXiv: 1003.5209 [quant-ph] (2010)

17. Rovelli, C.: Relational quantum mechanics. International Journal of Theoretical Physics 35, 1637-1657 (1996)

18. Yukalov, V.I. and Sornette, D.: How brains make decisions. Springer Proceedings in Physics 150, 37-53 (2014)

19. Helland, I.S.: The conception of God as seen from research on the foundation of quantum theory. Dialogo 4, 259-267 (2018)

20. Helland, I.S.: On religious faith, Christianity, and the foundations of physics. European Journal of Theology and Philosophy 2 (1), 10-17 (2021)

21. Hardy, L.:  Quantum theory from five reasonable axioms. arXiv: 0101012 [quant-ph] (2001)

22. Khrennikov, A.: Ubiquitous quantum structure: from psychology to finances, Springer, Berlin-Heidelberg-New
York (2010)

23. Busemeyer, J.R. and Bruza, P. D.: Quantum models of cognition and decision, Cambridge University Press, Cambridge (2012)

24. Aerts, D. and Beltran, L.: Are words the quanta of human language? Extending the domain of quantum cognition. arXiv:2110.04913 (2021)

25. Helland, I.S.: Simple counterexamples against the conditionality principle. The American Statistician 49, 351-356. Discussion 50, 382-386 (1995)

26. Smolin, L.: The Trouble with Physics. Houghton Mifflin, Boston (2007)

27. Rovelli, C.: Reality is Not What it Seems. Riverhead Books, New York (2017)

28. Susskind, L. and Lindsay, J.: An Introduction to Black Holes, Information and the String Theory Revolution. World Scientific, New Jersey (2005)

29. Laudal, O.A.: Mathematical Models in Science. World Scientific, New Jersey (2021)

30. Hardy, L.: Operational general relativity: possibilistic, probabilistic, and quantum. arXiv: 1608.06940 [gr-qc] (2016)

31. Mandl, F. and Shaw, G.: Quantum Field Theory. Chichester, West Surrex (2010)

32. Smolin, L.: The dynamics of difference. arXiv: 1712.04799v3 [gr-qc] (2019)

33. Smolin, L.: A real ensemble interpretation of quantum mechanics. Foundations of Physics 42, 1239-1261 (2012)

34. Smolin, L.: Quantum mechanics and the principle of maximum variety. Foundations of Physics 46, 736-758 (2015)

\end{document}